\documentclass[preprint,12pt]{elsarticle}

\usepackage{graphics}
\usepackage{graphicx}
\usepackage{epsfig,verbatim}
\usepackage{dcolumn}
\usepackage{amssymb}
\usepackage{amsmath}
\usepackage[english]{babel}
\usepackage[latin1]{inputenc}

\def\bea{\begin{eqnarray}}
\def\eea{\end{eqnarray}}
\newbox\grsign \setbox\grsign=\hbox{$>$} \newdimen\grdimen \grdimen=\ht\grsign
\newbox\simlessbox \newbox\simgreatbox
\setbox\simgreatbox=\hbox{\raise.5ex\hbox{$>$}\llap
     {\lower.5ex\hbox{$\sim$}}}\ht1=\grdimen\dp1=0pt
\setbox\simlessbox=\hbox{\raise.5ex\hbox{$<$}\llap
     {\lower.5ex\hbox{$\sim$}}}\ht2=\grdimen\dp2=0pt

\begin{document}

\begin{frontmatter}

\title{Langevin Simulation of Scalar Fields: \\ Additive and Multiplicative
  Noises and Lattice Renormalization}

\author[ift]{N. C. Cassol-Seewald}
\ead{nadiane@ift.unesp.br}
\author[ufsj]{R. L. S. Farias}
\ead{ricardofarias@ufsj.edu.br}
\author[ufrj]{E. S.  Fraga}
\ead{fraga@if.ufrj.br}
\author[ift]{G.  Krein}
\ead{gkrein@ift.unesp.br}
 \author[uerj]{Rudnei~O.~Ramos}
\ead{rudnei@uerj.br}
\address[ift]{Instituto de F\'{\i}sica Te\'orica, Universidade Estadual
  Paulista, \\ Rua Dr. Bento Teobaldo Ferraz, 271 - Bloco II, 
  01140-070 S\~ao Paulo, SP, Brazil}
\address[ufsj]{Departamento de Ci\^{e}ncias Naturais, Universidade Federal de
  S\~ao Jo\~ao Del Rei, \\
  36301-000 S\~ao Jo\~ao Del Rei, MG, Brazil}
\address[ufrj]{Instituto de
  F\'{\i}sica, Universidade Federal do Rio de Janeiro, \\
  21941-972 Rio de Janeiro, RJ, Brazil}
\address[uerj]{Departamento de
  F\'{\i}sica Te\'orica, Universidade do Estado do Rio de Janeiro, \\ 20550-013
  Rio de Janeiro, RJ, Brazil}

\begin{abstract}
We consider the Langevin lattice dynamics for a 
spontaneously broken $\lambda\phi^4$ scalar field 
theory where both additive and multiplicative noise 
terms are incorporated. The lattice renormalization 
for the corresponding stochastic Ginzburg-Landau-Langevin 
and the  subtleties related to the multiplicative 
noise are investigated. 
\end{abstract}

\end{frontmatter}

\section{Introduction}

The relevance of mathematical methods of quantum field theory for
statistical physics was recognized in the early seventies,
particularly in the investigation of equilibrium  phase transitions
and critical phenomena. Renormalization theory, originally linked with
the removal of infinities in perturbative calculations in quantum
field theory, turned out  to be a key element in the understanding of
critical phenomena \cite{zinn}. Functional integrals, Feynman  diagrams and loop
expansions are an integral part of the mathematical methods used
presently  in statistical physics -- Ref.~\cite{LeBellac:1991cq} is an
excellent introductory text on  concepts and methods of field theory
in the quantum and statistical domains. Likewise, a large class of
dynamic critical phenomena associated with time-dependent fluctuations
about equilibrium  states, naturally described in terms of stochastic
partial differential  equations~\cite{Hohenberg:1977ym,reviews}, can
also be formulated in terms of functional integrals  and therefore are
equivalently described by field theories~\cite{Martin:1973zz}. Among
the  variety of phenomena associated with the dynamics of phase
transitions, phase ordering~\cite{critical} seems  to be of particular
importance for the understanding of the time scales governing the
equilibration  of systems driven out of equilibrium. The influence of
the presence of an environment on the  dynamics of particles and
fields is encoded ``macroscopically'' in attributes that enter
stochastic evolution equations, usually in the form of dissipation and
noise terms. Relevant  time scales for different stages of phase
conversion can depend dramatically on the details  of these
attributes.  Noise terms, in particular, introduce difficulties in the
numerical  simulation of the evolution equations; difficulties related
to the well-known Rayleigh-Jeans  catastrophe in classical field
theory~\cite{parisibook}. 
The present paper is concerned with the use of field-theoretic
renormalization methods to control such catastrophe. 

A typical dynamical equation describing phase conversion is of the
form  of a Ginzburg-Landau-Langevin (GLL) equation~\cite{onuki}:

\begin{equation}
\tau \frac{\partial^2 \varphi} {\partial t^2} - \nabla^2 \varphi +  \eta \,
\dot\varphi + {\mathcal V}'_{\rm{eff}}(\varphi)=\xi({\bf{x}},t) \, ,
\label{langevin-simple}
\end{equation}
where $\varphi$ is a real scalar field, function of time and space
variables, ${\mathcal V}'_{\rm{eff}}(\varphi)$  is the field derivative of a
Ginzburg-Landau effective potential and $\eta$, which can be seen as a
response coefficient that defines time scales for the system and
encodes the intensity of dissipation, is usually taken to be a
function of temperature only, $\eta=\eta (T)$. The function
$\xi({\bf{x}},t)$ represents a stochastic (noise) force, assumed
Gaussian and white, so that 

\bea \langle \xi({\bf{x}},t)\rangle_\xi = 0,  \hspace{1.0cm} \langle
\xi({\bf{x}},t)\xi({\bf{x}}',t') \rangle_\xi = 2\,\eta T
\delta({\bf{x}}-{\bf{x}}')\delta(t-t'), 
\label{white}
\eea
in conformity with the fluctuation-dissipation theorem. The condensate
$\langle\varphi\rangle_\xi$ plays the role of the order parameter, where
$\langle \cdots \rangle_\xi$ means average over noise
realizations. The second  order time derivative in
Eq.~(\ref{langevin-simple}) appears naturally in  relativistic field
theories~\cite{{KUNS-820},Gleiser:1993ea} or when causality  is
incorporated via memory functions~\cite{{Koide:2006vf},{CCK}} in the
otherwise purely diffusive, first-order evolution
equations~\cite{{Hohenberg:1977ym},{onuki}}.  However, a more
complete, microscopic field-theory description of nonequilibrium
dissipative dynamics~\cite{calzetta-hu} shows that the complete form
for the effective GLL equation of motion can lead to much more
complicated scenarios than the one described by
Eq.~(\ref{langevin-simple}), depending on the allowed interaction
terms involving~$\varphi$. In general, there will be nonlocal
(non-Markovian) dissipation and colored noise, as well as the
possibility of field-dependent (multiplicative) noise terms
$\sim\varphi~\xi$ accompanied by density-dependent dissipation terms. 
Another typical example is provided by stochastic Gross-Pitaevskii (SGP) equations ~\cite{calzeta1,stoch-GPE} that incorporate thermal and quantum
fluctuations of a Bose-Einstein condensate (BEC) on the  traditional
mean-field equation~\cite{GP}. Such a stochastic equation can  be
derived from a microscopic inter-atomic Hamiltonian via the
closed-time-path  Schwinger-Keldysh effective action
formalism~\cite{calzetta-hu}.  In fact, on very general physical
grounds, one expects that dissipation effects should depend on the
local density $\sim\varphi^2\dot\varphi$ and, accordingly, the noise term
should contain a multiplicative piece $\sim\varphi\xi$.  The typical term
coming from fluctuations in the equation of motion for $\varphi$ will be
a functional of the form~\cite{calzetta-hu}
\begin{equation}
{\cal F}[\varphi(x)]=\varphi(x) \int d^4 x'  \varphi^2(x') K_1(x,x') + \int d^4
x'  \varphi(x') K_2(x,x') ,
\end{equation}
where $K_1(x,x')$ and $K_2(x,x')$ are nonlocal kernels  expressed in
terms of retarded Green's functions and whose explicit form depends on
the detailed nature of the interactions involving~$\varphi$. Explicit
treatments for these nonlocal kernels show that under appropriate
conditions one is justified to express the effective equation of
motion for $\varphi$ in a local form (see e.g. Ref.~\cite{BMR} and
references therein).   The existence of these additional terms in the
GLL equation will, of course, play an important role in the dynamics
of the formation of condensates. For instance, it was shown
that the effects of multiplicative noise are rather significant in the
Kibble-Zurek scenario of defect formation in one spatial
dimension~\cite{Antunes:2005ap}.

Although in the literature there are many different approaches for
studying the nonequilibrium dynamics in field
theory~\cite{aarts,berges,vega},  the use of stochastic Langevin-like
equations of motion is still a simpler and more direct approach 
in many different contexts in statistical physics and field
theory in general. For example,   some of us have considered the
effects of dissipation in the scenario of explosive spinodal
decomposition: the rapid growth of unstable modes following a quench
into the two-phase region of the quantum chromodynamics (QCD) chiral 
transition in the simplest
fashion~\cite{FK1}. Using a phenomenological Langevin  description
inspired by microscopic nonequilibrium field theory
results~\cite{Gleiser:1993ea,Greiner:1996dx,dirk}, the time evolution
of the order parameter in a chiral effective
model~\cite{Scavenius:2001bb} was investigated. Real-time
$(3+1)$-dimensional lattice simulations for the behavior of the
inhomogeneous chiral fields were performed, and it was shown that the
effects of dissipation could be dramatic in spite of the very
conservative assumptions that were made. Later, analogous but
even  stronger effects were obtained in the case of the deconfining
transition of  $SU(2)$ pure gauge theories using the same approach
\cite{Fraga:2006cr,Fraga:2007gg}.  Recent work in these directions by
a different group can be found in
Refs.~\cite{{arXiv:1105.0622},{arXiv:1111.3771}}.

In the present paper we consider the nonequilibrium dynamics of the
formation of a condensate in a spontaneously broken $\lambda \varphi^4$
scalar field theory within an improved Langevin framework which
includes the effects of multiplicative noise and density-dependent
dissipation terms. The corresponding stochastic GLL equation can be thought
of as a generalization of the results
of Ref.~\cite{Gleiser:1993ea} to the case of broken symmetry. The time
evolution for the formation of the condensate, under the influence of
additive and multiplicative noise terms, is solved numerically on a
$(3+1)$-dimensional lattice.  Particular attention is devoted to the
renormalization of the stochastic GLL equation in order to obtain
lattice-independent equilibrium results.

The paper is organized as follows. In Section II the proper lattice
renormalization of the GLL, in order to achieve equilibrium solutions
that are independent of lattice spacing, is addressed.  In Sec. III
the question of time discretization for a GLL with multiplicative
noise is discussed.   In Sec. IV we show the results of our lattice
simulations to study the behavior of the condensate. Section VI
contains our conclusions and perspectives. 

\section{Stochastic GLL equations}
\label{gll_equation}

In our study, we consider an extended GLL equation, incorporating
additive {\it and} multiplicative noise terms. The time evolution of
the field $\varphi({\bf{x}},t)$ at each point in space, which will
determine the approach of the condensate $\langle\varphi\rangle$ to
its equilibrium value will be dictated by the following equation:

\begin{equation}
\frac{\partial^2 \varphi}{\partial t^2} - \nabla^2\varphi + \left(
\eta_1 ~\varphi^2 + \eta_2 \right) \frac{\partial \varphi}{\partial t}
+ {\mathcal V}'_{\rm{eff}}(\varphi)= \xi_1({\bf{x}},t)~\varphi +
\xi_2({\bf{x}},t) \; ,
\label{langevin}
\end{equation}
where the dissipation coefficients $\eta_1$ and $\eta_2$ 
can be seen as response
coefficients that define time scales for the system and encode the
intensity of dissipation. The functions $\xi_1({\bf{x}},t)$
and $\xi_2({\bf{x}},t)$ represent stochastic (noise) forces, assumed
Gaussian and white, as in Eq.~(\ref{white}).
The motivation for Eq.~(\ref{langevin}) stems largely from explicit microscopic derivations
of the effective equation of motion for nonequilibrium quantum fields, where
this type of equation emerges naturally. We refer the interested 
reader to Refs. \cite{Gleiser:1993ea,calzetta-hu} for more details.

\subsection{Lattice renormalization}
\label{sec:latt-renorm}

Analytic solutions of Eq.~(\ref{langevin}) are achievable only in very
special situations (for the case of zero spatial dimensions, see e.g.
Ref.~\cite{CPC}), like in a linear approximation to ${\mathcal
  V}_{\rm{eff}}(\varphi)$, usually valid only at short times.  Complete
solutions describing the evolution of the system to equilibrium can be
obtained only through extensive numerical simulations. In general,
numerical simulations are performed on a discrete spatial lattice of
finite length under periodic boundary conditions.  However, in
performing lattice simulations of Eq.~(\ref{langevin}), one should be
careful in preserving the lattice-spacing independence of the results,
especially when one is concerned with the behavior of the system in
the continuum limit. The equilibrium probability distribution for the
field configurations $\phi$ that are solutions of
Eq.~(\ref{langevin}) is $P_{eq}[\phi] = e^{-S[\phi]}$, where
$S[\phi]$ is the Euclidean action. The corresponding partition
function is given by the path integral

\begin{equation}
Z[\phi] = \int \mathcal{D}\phi \; e^{-S[\phi]} \,.
\end{equation}

The calculation of expectation values and correlation functions
of~$\phi$ with this partition function leads to ultraviolet
divergences. In the presence of thermal noise, short and long
wavelength modes are mixed during the dynamics, yielding an unphysical
lattice spacing sensitivity. Such a lattice spacing sensitivity is
also present in the numerical simulation of SGP equations~\cite{cockburn,cockburn2}. 
The issue of obtaining
robust results, as well as the correct ultraviolet behavior, in
performing Langevin dynamics was discussed by several
authors~\cite{Borrill:1996uq,bett1,Gagne,bett2,{Fraga:2005wd},Fraga:2006uj}.

The problem, which is not {\it a priori} evident in the Langevin
formulation, is related to the well-known Rayleigh-Jeans ultraviolet
catastrophe in classical field theory ~\cite{parisibook}.  The
dynamics dictated by Eq.~(\ref{langevin}) is classical, and is
ill-defined for large momenta.  These {\it a priori} lattice
divergences can be eliminated by renormalizing the potential
${\mathcal V}_{\rm{eff}}$ through the addition of appropriate counterterms
(notice that these divergences are completely unrelated to the usual
ones of the quantum theory).  Since the divergent terms are all
perturbative, one can identify the appropriate diagrams and subtract
their result computed within the {\it classical theory}. Since the theory in three-dimensions is
super-renormalizable, only a mass renormalization is required. 
We only require, then, a renormalization of the mass
parameter $m^2_0$ in ${\mathcal V}_{\rm{eff}}(\varphi)$.   Using such a
renormalized potential in Eq.~(\ref{langevin}) leads to equilibrium
solutions $\varphi$ that are independent of the lattice spacing as we
are going  to explicitly show below. In practice, this lattice
renormalization procedure  corresponds to adding finite-temperature
counterterms to the original potential  ${\mathcal
  V}_{\rm{eff}}(\varphi)$, which guarantees the correct short-wavelength
behavior of the discrete theory as was originally shown by the authors
of Ref.~\cite{Farakos1994xh} within the framework of dimensional
reduction in a different context.

In order to define the notation and the procedure to calculate the
loop expansion of the effective potential~\cite{Jackiw}, we first
consider the theory in the continuum and switch to the lattice when
calculating the loop integrals.

\subsection{The classical field effective potential
  in the continuum: loop corrections}
\label{subsec:continuum}

The calculation of the classical three-dimensional loop corrections to
the potential starts as usual by introducing a constant field
$\varphi$ through the transformation

\begin{equation}
\phi \rightarrow \phi + \varphi \, ,
\end{equation}
and considering the action

\begin{equation}
\hat S[\varphi;\phi] \equiv S[\varphi+\phi] - S[\varphi] - \int d^3x\,
\phi \, \frac{\partial S[\phi]}{\partial \phi}\Bigg|_{\phi=\varphi} .
\end{equation}
Next, the quadratic terms in $\hat S$ are collected in a ``free"
action $\hat S_2$, and the remaining terms in an interacting action,
$\hat S_I$. Then, the effective classical field potential is defined
by the expression

\begin{equation}
e^{- \beta V \, {\mathcal V}_{\rm eff}[\varphi]} = e^{-\beta V \,
  {\mathcal V}[\varphi]} \int \mathcal{D}\phi\,e^{-\hat
  S[\varphi;\phi]} \,,
\end{equation}
where $V$ is the three-dimensional volume. {}From this expression, one
obtains for ${\mathcal V}_{\rm{eff}}[\varphi]$

\begin{eqnarray}
{\mathcal V}_{\rm{eff}}[\varphi] &=& {\mathcal V}[\varphi] - \frac{1}{\beta
  V} \ln \int \mathcal{D}\phi \, e^{-\hat S_2[\varphi;\phi]}
\nonumber \\ &-& \frac{1}{\beta V}  \ln \langle e^{-\hat
  S_I[\varphi,\phi]} \rangle \,,
\label{Veff}
\end{eqnarray}
where

\begin{equation}
\langle e^{- \hat S_I[\varphi,\phi]} \rangle = \frac{\int
  \mathcal{D}\phi\,e^{-\hat S_2[\varphi;\phi]} \, e^{- \hat
    S_I[\varphi;\phi]}} {\int \mathcal{D}\phi\,e^{-\hat
    S_2[\varphi;\phi]} } \,.
\end{equation}
The final step is the determination of $\varphi$ through

\begin{equation}
\frac{ d\,{\mathcal V}_{\rm{eff}}[\varphi] }{d\,{\mathcal \varphi}} = 0 \,.
\label{det-varphi}
\end{equation}

For a bare potential of the form ${\cal V}=-m_0^2\phi^2/2 + \lambda \phi^4/4!$, 
the action $\hat S_2[\varphi; \phi]$ is given by

\begin{equation}
\hat S_2[\varphi; \phi] = \beta \int d^3x \, \left[ - \frac{1}{2}
  \phi\nabla^2\phi + \frac{1}{2} m^2 \phi^2 \right] \,,
\end{equation}
where

\begin{equation}
m^2 = - m^2_0 + \frac{1}{2} \lambda \varphi^2 \,.
\label{m2}
\end{equation}
Since $\varphi$ is constant, the first functional integral in
Eq.~(\ref{Veff}) can be easily performed in momentum space, with the
result

\begin{equation}
\frac{1}{\beta V} \ln \int \mathcal{D}\phi \, e^{-\hat S_2[\varphi;
    \phi]} = - \frac{T}{2} \int_k \ln \widetilde G^{-1}[\varphi; k^2]
\,,
\end{equation}
where $\widetilde G^{-1}[\varphi; k^2]$ is the inverse of the
three-dimensional (classical field) propagator

\begin{equation}
\widetilde G[\varphi; k^2] = \frac{1}{k^2 + m^2} \,,
\end{equation}
and $\int_k \equiv \int \frac{d^3k}{(2\pi)^3}$.  The next divergent
contribution comes from the two-loop correction to the mass, 
which will require $\hat S_I$,

\begin{equation}
\hat S_I[\varphi; \phi] =  \beta \int d^3x \, \left( - \frac{1}{3}
\kappa \phi^3 + \frac{1}{4!} \lambda \phi^4\right) \,,
\end{equation}
with $\kappa = - \frac{1}{2} \lambda \varphi$.  Expansion of $e^{-\hat
  S_I}$ in Eq.~(\ref{Veff}) gives the two-loop divergent contribution,
that we will denote as $H[\varphi]$ and defined by

\begin{equation}
\frac{1}{\beta V} \ln \langle e^{- \hat S_I[\varphi,\phi]}
\rangle_{two-loop} = \frac{T^2}{2} \left(\frac{1}{3} \kappa
\right)^2 \, 6 \, H[\varphi] \, ,
\end{equation}
where

\begin{eqnarray}
H[\varphi] &=& \frac{1}{6V} \int d^3x d^3y \langle \phi^3(x) \phi^3(y)
\rangle \nonumber \\ &=& \int_k \int_q \, \widetilde G[\varphi; k^2]
\, \widetilde G[\varphi; q^2] \, \widetilde G[\varphi; (k+q)^2]
\,. \label{Hsun}
\end{eqnarray}
Now, the total divergent part of ${\mathcal V}_{\rm{eff}}[\varphi]$ is obtained
from

\begin{equation}
\frac{d^2{\mathcal
    V}_{\rm{eff}}[\varphi]}{d\varphi^2}\Bigg|_{\rm div. terms} = 
\frac{\lambda T}{2} \,  I_{div}[\varphi] -
\frac{\lambda^2 T^2 }{6} \, H_{div}[\varphi] , \label{divVeff}
\end{equation}
with

\begin{equation}
I[\varphi] =  \int_k \widetilde G[\varphi; k^2] \,, \label{Itad}
\end{equation}
where $I_{div}[\varphi]$ and $H_{div}[\varphi]$ represent the
divergent parts of $I[\varphi]$ and $H[\varphi]$. Notice that the
derivatives of $H[\varphi]$ with respect to $\varphi$ lead to finite
integrals and hence are irrelevant here. The evaluation of the
divergent parts of $I$ and $H$ requires a regularization scheme. Since
we simulate our GLL equations on a cubic lattice, we shall evaluate
these divergent parts by calculating the effective potential on the
lattice.

\subsection{The counterterms for the classical field
  potential on the lattice}
\label{subsec:lattice}

Here we consider the theory on a cubic lattice of volume $V = L^3$,
with $L=N a$, where $a$ is the lattice spacing and $N$ is the number
of lattice spacings. The coordinates $x_i$ of the lattice sites are
such that $0 \leq x_i \leq a(N-1)$. The Laplacian on the lattice,
$\nabla^2_{latt} \equiv \Delta$, is defined as

\begin{equation}
\Delta \phi(x) = \frac{1}{a^2} \sum^3_{i=1} \left[\phi(x+a\hat i) - 2
  \phi(x) + \phi(x-a\hat i)\right] \,,
\end{equation}
where $\hat i$ is the unit cartesian vector indicating the three
orthogonal directions of the square lattice. We also impose periodic
boundary conditions (PBC) on the fields,

\begin{equation}
\phi(x+aN\hat i) = \phi(x) \,,
\end{equation}
and define the {}Fourier transform $\tilde f(k)$ of a function $f(x)$
on the lattice as

\begin{equation}
\tilde f(k) = a^3 \sum_x e^{-ik{\cdot}x} f(x) \,.
\end{equation}
Because of the PBC, the allowed lattice momenta form the Brillouin
zone~${\mathcal B}$

\begin{equation}
k_i = \frac{2\pi}{aN} n_i \,, \hspace{0.5cm} n_i = 0,1,2,\cdots,N-1 .
\end{equation}
The inverse transform is given by

\begin{equation}
f(x) = \frac{1}{V}\sum_{k \, \in{\mathcal B}} e^{ik{\cdot}x} \tilde
f(k) \,,
\end{equation}
and the momentum summations over the Brillouin zone will be denoted by

\begin{equation}
\frac{1}{V}\sum_{k \, \in{\mathcal B}} \equiv \int_k \,.
\label{int_k}
\end{equation}

Let us now consider the derivation of the effective potential on the
lattice.  The lattice action is given by
\begin{equation}
S_{latt}[\phi] = a^3 \sum_x \, \left( - \frac{1}{2} \phi \Delta \phi +
{\mathcal V}[\phi] \right) \,, \label{Slatt}
\end{equation}
where ${\mathcal V}[\varphi]$ is the same as before.  The derivation
of the effective potential follows the same procedure as in the
continuum, leading to expressions for the one-loop and two-loop
divergent contributions as in Eqs.~(\ref{Hsun}) and (\ref{Itad}), 
but with $\int_k$
given by the sum over lattice momenta as indicated in
Eq.~(\ref{int_k}). What remains to be determined is the lattice
propagator corresponding to $\widetilde G$. The lattice propagator
$\widetilde G_{latt}$ can be obtained from the quadratic action $\hat
S_2[\varphi; \phi]$ on the lattice

\begin{equation}
\hat S_{2 \, latt}[\varphi; \phi] = a^3 \sum_x \, \left( - \frac{1}{2}
\phi \Delta \phi  + \frac{1}{2} m^2 \phi^2 \right) \,,
\end{equation}
with $m^2$ given by Eq.~(\ref{m2}). The lattice propagator
$G_{latt}[\varphi;x,y]$ is the inverse of $\left(-\Delta +
m^2\right)$, i.e.

\begin{equation}
a^3 \sum_y \left( - \Delta + m^2\right)_{x,y} G_{latt}[\varphi;x,y] =
\frac{1}{a^3} \delta_{x,y} \,,
\label{inv}
\end{equation}
where $\delta_{x,y}$ is the Kronecker delta. Since $\varphi$ is a
constant field, the solution of this equation can be obtained using
the Fourier transform of $G_{latt}[\varphi;x,y]$, defined as

\begin{equation}
G_{latt}[\varphi;x,y] = \int_k e^{ik{\cdot}x} \widetilde
G_{latt}[\varphi;k^2] \,.
\end{equation}
Substituting this into Eq.~(\ref{inv}), one obtains

\begin{equation}
\widetilde G_{latt}[\varphi,k^2] = \frac{a^2}{4}\frac{1}{d(\varphi;
  n_1,n_2,n_3)} \,,
\label{latt-prop}
\end{equation}
where

\begin{eqnarray}
d(\varphi; n_1,n_2,n_3) &=& \frac{a^2}{4}\left[ 2 a^{-2} \sum_{i=1}^3
  \left(1-\cos ak_i\right) + m^2 \right] \nonumber \\ &=&
\sum_{i=1}^{3} \sin^2(\pi n_i/N) + (am/2)^2 \,.
\end{eqnarray}
One can then write $I[\varphi]$ for the one-loop divergent term as

\begin{equation}
I[\varphi] =  \frac{1}{4aN^3} \sum^N_{n_i=0}
\frac{1}{d(\varphi;n_1,n_2,n_3)} \,,
\end{equation}
and the double sum in $H[\varphi]$ for the two-loop divergent term as

\begin{eqnarray}
H[\varphi] &=& \frac{1}{64N^6} \sum^{N-1}_{n_i,m_i=0}
\frac{1}{d(\varphi;n_1,n_2,n_3)d(\varphi;m_1,m_2,m_3)}
\nonumber\\ &\times& \frac{1}{d(\varphi;n_1+m_1,n_2+m_2,n_3+m_3)} \,.
\end{eqnarray}
The divergent parts of the sums above can be isolated in the limits of
$N \rightarrow \infty$ and $a \rightarrow 0$. For example, in the one-loop
term the 
sum in the limit of $N \rightarrow \infty$ can be converted into an
integral~\cite{Farakos1994xh},

\begin{eqnarray}
I[\varphi] &=& \frac{1}{4a\pi^3}\int^{\pi}_0 d^3x \, \frac{1}{\sum_i
  \sin^2x_i + (am/2)^2} \nonumber \\ &=& \frac{1}{4a} \int^\infty_0
d\alpha \, e^{-\alpha(am)^2/4}
\left[e^{-\alpha/2}I_0(\alpha/2)\right]^3,
\end{eqnarray}
where $I_0$ is the modified Bessel function. In the limit of $a
\rightarrow 0$, the divergent part of $I[\varphi]$ is given by

\begin{equation}
I_{div}[\varphi] = \frac{\Sigma}{4\pi a} \,,
\label{Idiv}
\end{equation}
where $\Sigma$ is a constant defined by

\begin{equation}
\Sigma = \frac{1}{\pi^2} \int_{-\pi/2}^{+\pi/2} d^3 x  \,
\frac{1}{\sum_i \sin^2x_i} \simeq 3.1759 \;.
\label{constSigma}
\end{equation}
It is important to remark here that for $T < T_c$ it is crucial the
use of Eq.~(\ref{det-varphi}) so that $m^2 < 0$ in the above
expressions. In a more demanding numerical computation,
$H_{div}[\varphi]$ can also be isolated, with the
result~\cite{Farakos1994xh}

\begin{equation}
H_{div}[\varphi] = \frac{1}{16\pi^2} \left[\ln \left(\frac{6}{a M}
  \right) + \zeta \right] \,,
\label{Hdiv}
\end{equation}
where $M$ is the renormalization scale and $\zeta \simeq 0.09$ is
another constant appearing in the integrals being evaluated
numerically.

The renormalized mass in the classical field perturbative loop
expansion is obtained as in the case of quantum field theory, i.e. by
subtracting the divergent parts of these graphs as indicated in
Eq.~(\ref{divVeff}):

\begin{equation}
- \frac{1}{2} \, m^2 \, \phi^2  \rightarrow  - \frac{1}{2} \, \left(
m^2 + \delta m^2 \right)  \, \phi^2  \equiv - \frac{1}{2} \, m^2_R \,
\phi^2 \,,
\end{equation}
where the mass counterterm is given by

\begin{equation}
\delta m^2 = \frac{\lambda T}{2} \; I_{div}[\varphi] - \frac{\lambda^2
  T^2 }{6} \; H_{div}[\varphi] \;.
\end{equation}
Therefore, we add the following finite-temperature counterterms to our
original potential:

\begin{equation}
\mathcal{V}_{ct}= \left\{ -\frac{ \lambda \Sigma}{8 \pi} \frac{T}{a}  +
\frac{\lambda^2}{96 \pi^2} T^2 \left[ \ln \left( \frac{6}{ a M}
  \right) + \zeta \right] \right\} \frac{\varphi^2}{2} \;.
\label{counterterms}
\end{equation}
Notice from the above equation that the dependence on the mass scale $M$
of the lattice counterterm is only logarithmic. So, it turns out that
results are only very weakly dependent on changes of the scale $M$. Of
course, any change in the renormalization scale $M$ can be compensated
by corresponding changes in the renormalized  parameters as also
expected from the renormalization group theory;  here only a change in
the coefficient of the term proportional to  $\varphi^2$ is required,
as it is clear from Eq. (\ref{counterterms}).

\section{Multiplicative Noise and Time Discretization}

The equation to be solved on the lattice is given by
Eq.~(\ref{langevin}).  We solve the GLL equation on the lattice with
periodic boundary conditions (PBC). When we insert the system in a
box, $\varphi$ acquires a discrete form $\varphi _{ijk}^{n}$, where
$t=n\Delta t$ with $n=0,1,2, \ldots$, $x=ia$, $y=ja$ and $z=ka$, 
$a$ being the lattice spacing $a=\frac{L}{N}$.
Using PBC we have

\begin{eqnarray}
\varphi _{N+1jk}^{n} &=&\varphi _{1jk}^{n}\text{ ; }\varphi _{iN+1k}^{n}=\varphi
_{i1k}^{n}\text{ ; }\varphi _{ijN+1}^{n}=\varphi _{ij1}^{n}\;, \nonumber
\\ \varphi _{N+1N+1k}^{n}&=&\varphi_{11k}^{n}\text{ ; }\varphi
_{N+1jN+1}^{n}=\varphi _{1j1}^{n} \text{ ; }  \varphi _{iN+1N+1}^{n} =\varphi
_{i11}^{n} \;, \nonumber \\ \varphi _{N+1N+1N+1}^{n} &=&\varphi
_{111}^{n}\text{ ; } \varphi _{0jk}^{n}=\varphi_{Njk}^{n}\text{ ; }\varphi
_{i0k}^{n}=\varphi _{iNk}^{n}  \;, \nonumber \\ \varphi _{ij0}^{n} &=&\varphi
_{ijN}^{n}\text{ ; }\varphi _{i00}^{n}=\varphi _{iNN}^{n} \text{ ; } \varphi
_{0j0}^{n} =\varphi _{NjN}^{n} \;, \nonumber \\ \varphi _{00k}^{n}&=&\varphi
_{NNk}^{n}\text{ ; } \varphi _{000}^{n}=\varphi _{NNN}^{n} \;.
\label{eqa1}
\end{eqnarray} 
We write the Laplacian as

\begin{eqnarray}
\nabla ^{2}\varphi _{ijk}^{n} &=&\frac{\partial ^{2}\varphi
  _{ijk}^{n}}{\partial ^{2}x}+\frac{\partial ^{2}\varphi
  _{ijk}^{n}}{\partial ^{2}y}+\frac{\partial ^{2}\varphi
  _{ijk}^{n}}{\partial ^{2}z}  \nonumber \\ &=&\frac{1}{a}\left[
  \left( \frac{\varphi _{i+1jk}^{n}-\varphi _{ijk}^{n}}{a} \right) -\left(
  \frac{\varphi _{ijk}^{n}-\varphi _{i-1jk}^{n}}{a}\right) \right.
  \nonumber \\ &+&\left( \frac{\varphi _{ij+1k}^{n}-\varphi
    _{ijk}^{n}}{a}\right) -\left( \frac{ \varphi _{ijk}^{n}-\varphi
    _{ij-1k}^{n}}{a}\right)  \nonumber \\ &+&\left. \left( \frac{\varphi
    _{ijk+1}^{n}-\varphi _{ijk}^{n}}{a}\right) -\left(  \frac{\varphi
    _{ijk}^{n}-\varphi _{ijk-1}^{n}}{a}\right) \right] \nonumber \\ &=&
\frac{1}{a^{2}}\left[ \varphi _{i+1jk}^{n}+\varphi _{ij+1k}^{n}+\varphi
  _{ijk+1}^{n}\right.  \nonumber \\ &-&\left. 6\varphi _{ijk}^{n}+\varphi
  _{i-1jk}^{n}+\varphi _{ij-1k}^{n}+\varphi _{ijk-1}^{n}\right] \;.
\label{eqa2}
\end{eqnarray}
{}For simplicity, we introduce the compact notation

\begin{equation}
\nabla ^{2}\varphi _{ijk}^{n}=\left( L\varphi \right) _{ijk}^{n}\;.
\end{equation}
In addition, we divide the time in $n$ steps, $t=n\Delta t$, with $
n=0,1,2,\ldots$. With respect to the discretization of the time
derivatives, we use the leapfrog approximation method, where the
algorithm is defined by the following iteration scheme:

\begin{eqnarray}
\frac{\partial \varphi _{n}}{\partial t}
&=&\dot{\varphi}_{n}=\frac{1}{2}\left(
\dot{\varphi}_{n+1/2}+\dot{\varphi}_{n-1/2}\right)\;,  \nonumber
\\ \dot{\varphi}_{n+1/2} &=&\frac{1}{\Delta t}\left( \varphi _{n+1}-\varphi
_{n}\right)  \;, \nonumber \\ \frac{\partial ^{2}\varphi _{n}}{\partial
  t^{2}} &=&\ddot{\varphi}_{n}=\frac{1}{ \Delta t}\left(
\dot{\varphi}_{n+1/2}-\dot{\varphi}_{n-1/2}\right)\;.
\end{eqnarray}
Now we rewrite (\ref{langevin}) in terms of discrete the field $\varphi
_{ijk}^{n}$	

\begin{eqnarray}
\frac{\partial^{2}\varphi _{ijk}^{n}}{\partial t^{2}}-\nabla^{2}\varphi
_{ijk}^{n}+\eta_{1}\left( \varphi_{ijk}^{n}\right)^{2}\frac{\partial \varphi
  _{ijk}^{n}}{\partial t}+\eta _{2}\frac{\partial
  \varphi_{ijk}^{n}}{\partial t} +{\cal V}_{\rm eff}^{\prime }\left(
\varphi_{ijk}^{n}\right)\nonumber\\
 = \varphi_{ijk}^{n}\xi_{1}+\xi_{2} \;.
\end{eqnarray}
Using the discretized quantities derived above we obtain

\begin{eqnarray}
\frac{1}{\Delta t}\left( \dot{\varphi}_{ijk}^{n+1/2}-\dot{\varphi}
_{ijk}^{n-1/2}\right)  &=&\left( L\varphi \right) _{ijk}^{n}-\eta
_{1}\left( \varphi _{ijk}^{n}\right) ^{2}  \frac{1}{2}\left(
\dot{\varphi}_{ijk}^{n+1/2}+\dot{\varphi} _{ijk}^{n-1/2}\right)  \nonumber
\\ &-&\eta _{2}\frac{1}{2}\left( \dot{\varphi}_{ijk}^{n+1/2}+\dot{\varphi}
_{ijk}^{n-1/2}\right)  \nonumber \\ &-&{\cal V}_{\rm eff}^{\prime
}\left( \varphi _{ijk}^{n}\right)  +\varphi_{ijk}^{n}\xi _{1}+\xi _{2} \;,
\end{eqnarray}
The equation that we solve iteratively is then

\begin{eqnarray}
&&\left\{ 1+\frac{1}{2}\left[ \eta _{1}\left( \varphi _{ijk}^{n}\right)
  ^{2}+\eta _{2}\right] \Delta t\right\}  \dot{\varphi}_{ijk}^{n+1/2} \nonumber\\
&=&
\left\{ 1-\frac{1}{2}\left[ \eta _{1}\left( \varphi _{ijk}^{n}\right)
  ^{2}+\eta_{2}\right] \Delta t\right\} \dot{\varphi} _{ijk}^{n-1/2} +
\left( L\varphi \right)_{ijk}^{n}\Delta t - {\cal V}_{\rm
  eff}^{\prime }\left( \varphi _{ijk}^{n}\right) \Delta t \nonumber \\&+& \varphi
_{ijk}^{n}\xi _{1}\Delta t+\xi _{2}\Delta t \;.
\end{eqnarray}
In a compact notation we have

\begin{eqnarray}
\dot{\varphi}_{ijk}^{n+1/2} &=&\frac{1}{\Xi }\left[ \dot{\varphi}
  _{ijk}^{n-1/2}\Theta +\left( L\varphi \right) _{ijk}^{n}\Delta t- {\cal
    V}_{\rm eff}^{\prime}\left( \varphi _{ijk}^{n}\right) \Delta t\right.
  \nonumber \\ &+&\left. \varphi _{ijk}^{n}\xi _{1}\Delta t+\xi
  _{2}\Delta t\right] \;,
\label{leapfrog}
\end{eqnarray}
where

\begin{eqnarray}
\Xi &=&1+\frac{1}{2}\left[ \eta _{1}\left( \varphi _{ijk}^{n}\right)
  ^{2}+\eta _{2}\right] \Delta t \;, \nonumber \\ \Theta
&=&1-\frac{1}{2}\left[ \eta _{1}\left( \varphi _{ijk}^{n}\right)
  ^{2}+\eta _{2}\right] \Delta t\;.
\end{eqnarray}

To update the field we make:

\begin{equation}
\varphi_{ijk}^{n+1}=\varphi_{ijk}^{n}+\Delta t\, \dot{\varphi}_{ijk}^{n+1/2}
\;.
\end{equation}

On the lattice and with discretized time, the noise terms are modeled to
satisfy the discretized fluctuation-dissipation relation:

\begin{equation}
\langle \xi_{i,n} \xi_{j,n'} \rangle = 2 \eta_i T \delta_{i,j}
\delta_{n,n'}/(a^3 \Delta t)\;,
\end{equation}
with an amplitude that can then be written as

\begin{equation}
\xi_{i,n} = \sqrt{ \frac{2 \eta_i T}{a^3 \Delta t}} G_{i,n}\;,
\label{xii}
\end{equation}
where $G_{i,n}$ is obtained from a zero-mean unit-variance Gaussian.
Here, indices $i=1,2$ denote one of the two noises and respective
dissipation terms  appearing in the generalized GLL equation,
Eq.~(\ref{langevin}).

{}Finally, since we are dealing with multiplicative noise in a
Langevin equation, there is a well-known ambiguity in the
discretization of time \cite{Arnold:1999uza}. This ambiguity is
manifest whenever there are multiplicative noises and they are
delta-correlated in time (i.e.,  they are Markovian, like in the
approximation we have adopted in our simulations), and leads to the
two most common types of discretization procedures, the one  by Ito
and the one by Stratonovich~\cite{gardiner}. The leapfrog scheme shown
in Eq.  (\ref{leapfrog}) corresponds to Ito's prescription. The
ambiguity can be seen when one discretizes the (Markovian)
multiplicative noise term. Notice from Eq. (\ref{xii}) that the noise
terms in Eq. (\ref{leapfrog}) are actually of order $(\Delta
t)^{1/2}$, instead of order $\Delta t$ as the remaining terms.  So,
the multiplicative noise term needs to be re-expanded up the to next
order in $\Delta t$.  This can be performed when one writes the
discretized form for the multiplicative noise term using the Riemann
formula,

\begin{equation}
\int_t^{t + \Delta t} \varphi({\bf x}, t') \xi_1({\bf x}, t') dt' =
\left[(1-\alpha) \varphi({\bf x}, t) + \alpha \varphi({\bf x}, t+ \Delta t)
  \right] \chi_1 ({\bf x}, t)\;,
\label{riemann}
\end{equation}
where $0 \leq \alpha \leq 1$, with the Stratonovich prescription
corresponding to $\alpha=1/2$, while Ito's corresponds to 
$\alpha=0$ \cite{gardiner}. In Eq.~(\ref{riemann}) $\chi_1({\bf x},t)$ 
is a new Gaussian stochastic process described by (in discretized form)

\begin{equation}
\langle \chi_{1,n} \chi_{1,n'} \rangle = 2 \eta_1 T\delta_{n,n'}
\Delta t/a^3\;.
\end{equation}

Using Eq.~(\ref{riemann}) back in the leapfrog equation,
Eq.~(\ref{leapfrog}), and re-expanding it to the next order in the
multiplicative noise term, one finds  that in the Stratonovich
interpretation there is an additional correction term to the leapfrog
equation in Eq.~(\ref{leapfrog}) of order $\alpha
\dot{\varphi}_{ijk}^{n-1/2} \chi_{1}\Delta t \Theta/\Xi^2$, which is
already an order $(\Delta t)^{1/2}$ higher than the remaining terms in
that equation. We have explicitly checked in all our simulations that
this is a negligible correction (which is
consistent with the stochastic second order differential
equations in time discussed in \cite{Arnold:1999uza}) for all our results, 
and so we have
adopted Ito's prescription, i.e. Eq. (\ref{leapfrog}). Notice that
this might not be the case would we be working with first order in
time derivative equations, in which case large corrections due to the
difference between  Ito's and Stratonovich's interpretations can arise
\cite{ojalvo}.  {}For a discussion on different prescriptions
in the context of the relativistic Brownian motion, see
e.g. Ref. \cite{Koide:2007sv}.

\section{Numerical Results}

In this Section we present our results for the numerical simulations
of the stochastic GLL equation in three spatial dimensions according to
the discretization method described in the previous section. 

Our interest is in the relaxation of a field to its equilibrium
configuration. In order to analyze this behavior, we start by studying
the dependence of the solutions on the lattice spacing $a = L/N$ and
then, by introducing the lattice counterterms discussed and derived in
the previous section, we show how lattice-independent results can be
obtained in Langevin simulations using both standard additive noise
and dissipation terms and also in its generalized form, which includes
field-dependent (multiplicative) noise and density-dependent
dissipation. This is particularly important, since as we have discussed
in the Introduction, the effective equations of motion for background
scalar fields turn out to be in general of the generalized Langevin
form.

All results that will be presented here refer to the time dependence
of the volume average of the noise-averaged order parameter, defined
as

\begin{equation}
\langle \varphi(x,y,z,t) \rangle = \frac{1}{N^3} \sum_{i j k}
\bar{\varphi}^n_{ijk} \;,
\label{campo1}
\end{equation}
where $\bar{\varphi}^n_{ijk}$ is the average over a large number $N_r$ of
independent noise realizations:

\begin{equation}
\bar{\varphi}^n_{ijk}= \frac{1}{N_r} \sum_{r=1}^{N_r}\varphi_{ijk}^{n} \;.
\label{campo2}
\end{equation}

In all our numerical Langevin simulations, we consider $N_r$ between
20 and 100. As usual, lattices with larger values of $L$ require
relatively less realizations over the noise.  We have considered and
tested different lattice sizes to ensure the robustness of all
numerical results.

\subsection{The problem of lattice dependence in the generalized GLL approach}

\begin{figure}[htb]
  \vspace{0.7cm}
  \centerline{\epsfig{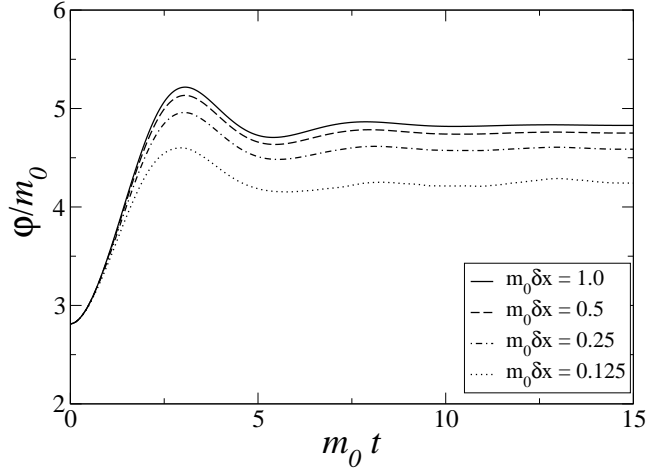}}
\caption[]{Solution of the standard GLL equation
  (\ref{langevin-simple})  using the leap frog algorithm for different
  lattice spacings.
  \label{noCTa}}
\end{figure}

\begin{figure}[htb]
  \vspace{0.7cm}
  \centerline{\epsfig{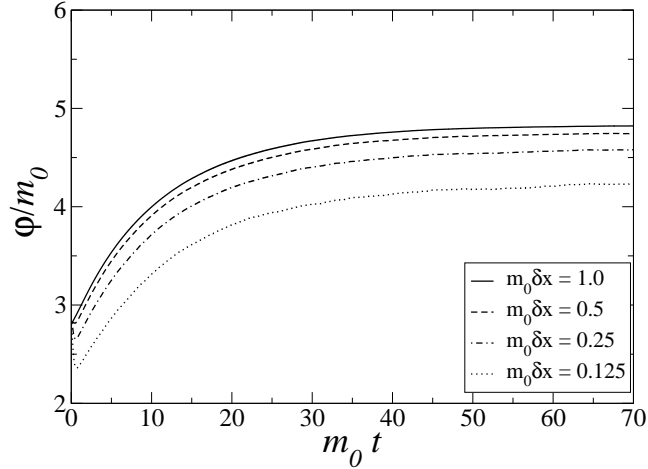}}
\caption[]{Solution of the GLL equation with both additive and
  multiplicative noise and dissipation terms using the leap frog
  algorithm for different lattice spacings. The parameters are
  the same as in {}Fig. \ref{noCTa}.
  \label{noCTam}}
\end{figure}

As discussed in the previous section, the simulation of equations with
noise, being classical by nature, leads to the appearance of
Rayleigh-Jeans ultraviolet divergences at long times when simulating
the equation on a discrete lattice. These divergences manifest
themselves in the form of lattice-spacing dependence of the
equilibrium solutions.  We can show that by just considering the
easiest nonequilibrium evolution, which is the one of relaxation to
the equilibrium state with initial conditions away from it.  In
{}Figs.  \ref{noCTa} and \ref{noCTam} we show the corresponding
dynamics for the scalar field expectation value, $\langle
\varphi(x,y,z,t) \rangle$ defined in Eqs.~(\ref{campo1}) and
(\ref{campo2}), for a symmetry-broken Ginzburg-Landau
quartic potential, defined as

\begin{equation}
\mathcal{V}(\varphi)= \frac{\lambda}{24}\left(T^2 - T_c^2\right) \frac{\varphi^2}{2} 
+ \frac{\lambda}{4 !}\varphi^4\;,
\label{Veff-GL}
\end{equation} 
for $T<T_c$, with the critical temperature $T_c^2 = 24 m_0^2/\lambda$ 
extracted from the finite-temperature effective potential.

The initial state for the field in the simulations for the broken
phase was taken around the inflexion (or spinodal) point of the
finite temperature (Ginzburg-Landau) potential, $\varphi_{\rm{infl}}$, defined by

\begin{equation}
\frac{d^2 \mathcal{V}(\varphi,T)}{d \varphi^2}\Bigr|_{\varphi = \varphi_{\rm
    infl}} =0\;.
\label{phiinfl}
\end{equation}

In {}Fig. \ref{noCTa} we show results for the standard Langevin
equation, Eq.~(\ref{langevin-simple}) with $\tau = 1$, while
{}Fig. \ref{noCTam} displays  results for the generalized case,
Eq.~(\ref{langevin}). The parameter values considered  for the
temperature and dissipation terms $\eta_1$ and $\eta_2$, in units of
$m_0$, were  $T/m_0 = m_0 \eta_1 = \eta_2/m_0  \equiv 1$, while the
dimensionless quartic coupling constant was chosen to be
$\lambda=0.25$. The scale $M$ is taken as $M/m_0=1$. These values
suffice for our purposes of just demonstrating the lattice dependence
problem in Langevin simulations. In {}Figs. \ref{noCTa} and
\ref{noCTam} the number of lattice points and the time stepsize were
kept constant, $N=64$ and $\delta t=0.01$ (in units of $m_0$),
respectively, while the lattice spacing, $m_0\delta x \equiv a=L/N$, was
varied.

It is clear that the solutions shown in {}Figs. \ref{noCTa} and
\ref{noCTam} are not stable as the lattice spacing is modified.  As
discussed previously, this problem can be traced to the fact that the
equilibrium value of the quantity $\bar{\varphi}_{ijk}$ gives the
classical average

\begin{eqnarray}
\bar{\varphi}_{ijk} =\frac{\int \emph{D}\varphi \, \varphi_{ijk} \, e^{-\beta
    \mathcal{V}\left( \varphi \right)}}{\int \emph{D}\varphi \,\, e^{-\beta \mathcal{V}\left(
    \varphi \right) }} \;,
\end{eqnarray}
a divergent quantity. Thus, stable equilibrium solutions of the GLL
equation, i.e., solutions not sensitive to lattice spacing, can only
be obtained by the introduction of the appropriate counterterms in the
effective potential in order to eliminate these divergences. These
counterterms are the ones derived in the previous section and given by
Eq.~(\ref{counterterms}).

\begin{figure}[htb]
  \vspace{0.7cm}
  \centerline{\epsfig{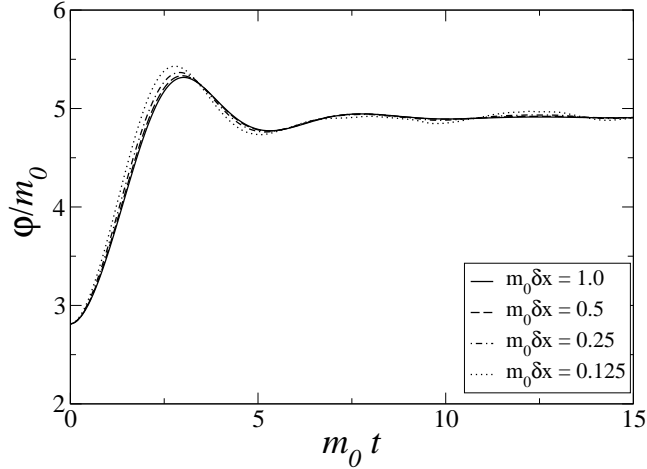}}
\caption[]{Solution of the standard GLL equation using the leap  frog
  algorithm for different lattice spacings and including the
  renormalization counterterms.
  \label{CTa}}
\end{figure}

\begin{figure}[ht]
  \vspace{1cm}
  \centerline{\epsfig{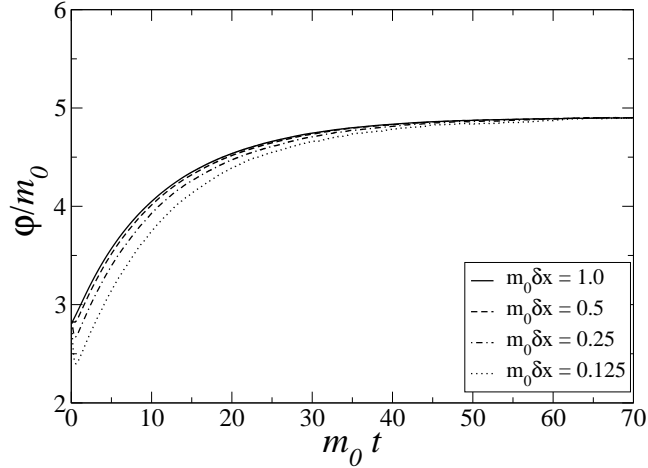}}
\caption[]{Solution of the GLL equation with both additive and
  multiplicative noise, and dissipation terms and including the
  renormalization counterterms.
  \label{CTam}}
\end{figure}

In {}Figs. \ref{CTa} and \ref{CTam} we present results of the
simulations including the counterterms. As we can see, equilibrium
solutions that are independent of lattice spacing are obtained.

{}For the standard GLL equation this was shown extensively in a series
of previous papers~\cite{Borrill:1996uq,bett1,Gagne,bett2}, but we are
not aware of the same demonstration for the case including
multiplicative noise terms.  {}From the results shown in
{}Figs. \ref{CTa} and \ref{CTam} we can also immediately draw a couple
of interesting conclusions.  The first is that even though the
counterterms used were calculated with an equilibrium partition
function, thus ensuring lattice-spacing independence only in
equilibrium (large-time) situations, the results for short times show
only small lattice-spacing dependence. Second, we can notice that the
relaxation time scales to the equilibrium state is about the same in
all cases.  Another observation we can make based on the results shown
in {}Figs. \ref{CTa} and \ref{CTam} is that the dynamics with
multiplicative noise and dissipation terms can be quite different from
that driven by additive noise only. In particular, notice from
{}Fig. \ref{CTam} that the relaxation time to the equilibrium state
with the generalized GLL equation is much longer than the one with
just additive noise, even though the magnitudes of the dissipation
coefficients (in the dimensionless units in terms of $m_0$) are of order
one.   One also notices that the difference in the dynamics of the two cases 
shown in {}Figs. \ref{CTa} and \ref{CTam}, 
and in special the overdamped behavior seen in {}Fig. \ref{CTam}, 
is just a consequence of the parameters chosen. 
In particular, since in the multiplicative 
noise case the dissipation term is proportional to the square of the
amplitude of the field, $\varphi^2$, and, for the parameters chosen,
$\varphi \sim 2.5 - 5$, this corresponds to a much more 
intense dissipation than the one for the case of additive noise, 
thus explaining the difference in the dynamics.

\section{Conclusions}

In this work we have studied several important aspects regarding the
dynamics of a scalar field background configuration in broken phase. 
It has been long recognized that the
effective evolution equation for the field can be of a complicated
form. {}From the analogy with standard Langevin equations for the
study of the approach to equilibrium, the microscopically derived
effective evolution equation allows for the presence of similar
additive noise and dissipation terms, but also for multiplicative
(field-dependent) noise and dissipation contributions. Although equations of
motion of the standard Langevin form (with only additive noise and
dissipation) have been extensively studied in the literature, its
generalized form, which includes the multiplicative noise and
dissipation terms, still demands extensive studies.  Here we have
performed a number of numerical simulations with these equations on a
cubic lattice. We have also pointed out another issue
frequently overlooked in the literature: the necessity of adding
lattice renormalization counterterms to cancel Rayleigh-Jeans
divergences of the corresponding classical theory in order to produce
sensible results from Langevin simulations.  We have shown that the
same lattice counterterms that are required for the standard Langevin
simulations also work for the generalized Langevin equations,
producing lattice-independent equilibrium quantities, and also
minimizing the dependence of the dynamics on the lattice parameters.
One must also notice that the evaluation of correlation functions
will also need appropriate counterterms to render the results lattice
independent. {}For example, a quadratic correlation like $\langle \varphi^2 \rangle$
will require a counterterm that can be expressed in terms of $I_{\rm div}$, 
Eq. (\ref{Idiv}).
A cubic correlation like $\langle \varphi^3 \rangle$ will require an
additional counterterm that is given in terms of 
$H_{\rm div}$, Eq. (\ref{Hdiv}). Higher-order
correlations are expected to be free of divergences (i.e., will not require
counterterms). That only these two terms, the ones which render
the classical effective potential finite, are necessary could be anticipated 
by recalling that the effective potential is the generator of (zero-momentum)
one-particle irreducible Green's functions. Thus, higher-order Green's functions
should not require other types of counterterms.

One interesting and important problem that still remains, though
beyond the objectives set for the present work, is the study of the
validity of the approximation of transforming the complicated nonlocal
(non-Markovian) equations of motion obtained through a microscopic
derivation (via the Schwinger-Keldysh real-time formalism for the
effective action) into the local form used in the simulations
performed for this study. This is a notoriously difficult problem due
to the oscillatory nature of the nonlocal kernels appearing in the
full effective equation of motion, which leads to uncontrollable
numerical behavior in simulations. Simulations and studies of the effects of the
nonlocal terms in zero spatial dimensions~\cite{PRE,nonmarkov,hasegawa1} have given
indications of the importance of the full non-Markovian  dynamics
compared with their local approximation. Implementing  the space
dependence on the non-local kernels and a full simulation of the
non-Markovian dynamics is the subject of a future work that is expected
to complement the present investigation.

\section{Acknowledgements}
E.S.F. would like to thank T. Kodama, T. Koide, A. J. Mizher and L. F. Palhares for discussions 
on related matters. This work was partially supported by CAPES, CNPq, FAPERJ, FAPESP, FAPEMIG and
FUJB/UFRJ (Brazilian Agencies).

\end{document}